# A new method using deep transfer learning on ECG to predict the response to cardiac resynchronization therapy

Zhuo He[a], Hongjin Si[b, c], Xinwei Zhang[b], Qing-Hui Chen[d], Jiangang Zou[b*], Weihua Zhou[a, e*]

[a] Department of Applied Computing, Michigan Technological University, Houghton, MI, USA
[b] Department of Cardiology, The First Affiliated Hospital of Nanjing Medical University, Nanjing, Jiangsu, China
[c] Department of Cardiology, The Affiliated Huaian No.1 People's Hospital of Nanjing Medical University, Huaian, Jiangsu, China
[d] Department of Kinesiology and Integrative Physiology, Houghton, MI, USA
[e] Center for Biocomputing and Digital Health, Institute of Computing and Cybersystems, Health Research Institute, Michigan Technological University, Houghton, MI

*Address for correspondence*

Weihua Zhou

E-mail: whzhou@mtu.edu

Address: 1400 Townsend Drive, Houghton, MI 49931, USA

Or

Jiangang Zou

E-mail: jgzou@njmu.edu.cn

Address: Guangzhou Road 300, Nanjing, Jiangsu, China 210029

*Running Head*

DL on ECG to predict CRT response

*Acknowledgements*

This research was supported by a Michigan Technological University Research Excellence Fund Research Seed grant (PI: Weihua Zhou) and a research seed grant from Michigan Technological University Health Research Institute (PI: Weihua Zhou). This research was also supported by the National Natural Science Foundation of China

*Abstract*

**Background**: Cardiac resynchronization therapy (CRT) has emerged as an effective treatment for heart failure patients with electrical dyssynchrony. However, accurately predicting which patients will respond to CRT remains a challenge. This study explores the application of deep transfer learning techniques to train a predictive model for CRT response.
**Methods**: In this study, the short-time Fourier transform (STFT) technique was employed to transform ECG signals into two-dimensional images. A transfer learning approach was then applied on the MIT-BIT ECG database to pre-train a convolutional neural network (CNN) model. The model was fine-tuned to extract relevant features from the ECG images, and then tested on our dataset of CRT patients to predict their response.
**Results**: Seventy-one CRT patients were enrolled in this study. The transfer learning model achieved an accuracy of 72% in distinguishing responders from non-responders in the local dataset. Furthermore, the model showed good sensitivity (0.78) and specificity (0.79) in identifying CRT responders. The performance of our model outperformed clinic guidelines and traditional machine learning approaches.
**Conclusion**: The utilization of ECG images as input and leveraging the power of transfer learning allows for improved accuracy in identifying CRT responders. This approach offers potential for enhancing patient selection and improving outcomes of CRT.

*Key words*
Cardiac resynchronization therapy, ECG, deep learning, convolutional neural network, transfer learning

# Introduction

The current guidelines for patient selection of cardiac resynchronization treatment (CRT) primarily rely on ECG-based criteria, namely QRS duration and morphology [1,2]. While QRS duration has demonstrated clinical value in predicting CRT response, it lacks the finesse to predict response on a patient-specific level accurately [1,3]. Multiple QRS cutoff values have been considered in different trials and studies, and the electrical LBBB pattern is widely accepted as a strong predictor of CRT response. Among LBBB patients, CRT response improves as QRS duration increases; on the other hand, the benefit of CRT starts to emerge in non-LBBB patients when QRS duration is ≥ 150ms [1,3–5]. However, ECG is not always effective in measuring the presence or severity of electrical dyssynchrony in all ventricular segments; and only significant myocardial masses can affect QRS morphology and duration [6]. Moreover, QRS duration alone is not specific enough to characterize exact electrical patterns [7–9]. Even in cases of LBBB, different and heterogeneous electrical activation patterns can exist despite similar QRS morphology and duration [10,11]. Therefore, researchers are seeking more accurate and patient-specific predictors beyond QRS duration and morphology.

Transfer learning is a powerful technique that leverages empirical knowledge gained from solving one problem to solve a related but different problem. In medical research, transfer learning has been widely adopted due to the limited availability of annotated medical images and the high cost of obtaining annotations [12]. The general process of transfer learning involves pre-training a deep neural network on a large dataset and then fine-tuning the network on a smaller target dataset. This approach enables the transfer of knowledge learned from the source dataset to the target dataset, resulting in improved performance with fewer data [13]. This allows the data of arrhythmia patients in large public databases to be better generalized and used for CRT patient selection.

In this study, we first represent the ECG signals obtained from the MIT-BIH arrhythmia dataset [14] with two-dimensional images by the short-time Fourier transform (STFT) technique, using a transfer learning approach to extract the input image features from the convolutional neural network (CNN) model (ResNet). Next, we fine-tuned the pre-trained models to extract features from the ECG of CRT data. We used them as inputs to classifiers such as logistic regression, support vector machine (SVM), and random forest (RF) to classify CRT patients based on ECG data, respectively.

## Methods

In this study, the 1-D ECG signal was reconstructed as a 2-D time-frequency spectrogram image for obtaining information on time, frequency, and energy of the heartbeats. Figure 1 shows the flowchart of the proposed method.

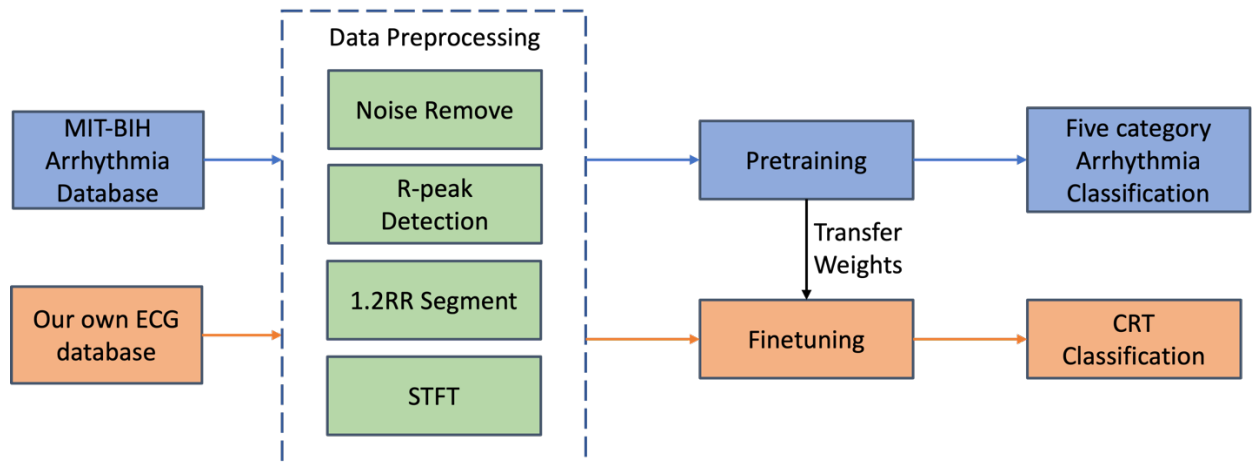

Figure 1. Visualization of transfer learning in this work. The process is divided into 4 steps: (1) Data preprocessing for both the MIT-BIH arrhythmia database and our own dataset to convert 1D ECG signals to 2D images (2) deep convolutional neural network (CNN) is pretrained on the MIT-BIH arrhythmia database for a selected pretraining objective, e.g. classification of arrhythmia; (3) the pretrained weights are used as initial weights of a new CNN; (4) this CNN is finetuned on our own database to predict CRT response.

## Data

The data used in this study is sourced from two separate databases. The first database is the MIT-BIH Arrhythmia Database, which contains over 109,000 annotated ECG recordings of 47 subjects at 360 Hz between 1975 and 1979 [14,15]. Heartbeats are annotated by two or more cardiologists independently. Fourteen original heartbeat types are consolidated into five groups according to the Association for the Advancement of Medical Instrumentation (AAMI) recommendation. This database is widely used in the research community and is considered a benchmark for arrhythmia detection algorithms. The data from the MIT-BIH database was used to pre-train the deep learning models for transfer learning. Specifically, the weights of a pre-trained model on the MIT-BIH dataset were used to initialize the weights of the model for training on our own dataset.

The second database used in this study was a cohort of 71 CRT patients from the First Affiliated Hospital of Nanjing Medical University. All the patients had an LV ejection fraction (LVEF) $\leq$ 35%, QRS duration > 120ms, New York Heart Association (NYHA) functional class II to IV symptoms, and optimal medical therapy at least 3 months before CRT. All the patients underwent resting gated SPECT MPI, echocardiography, and NYHA function classification at baseline and 6 months after CRT. This data was used to fine-tune and evaluate the performance of the deep learning models on new, unseen data. This study complied with the Declaration of Helsinki and was approved by local ethics committees. All patients gave written informed consent.

## Evaluation of LV function by echocardiography

Echocardiography data of all patients were assessed by experienced ultrasound experts blinded to any clinical data and MPI data before and 6 months after CRT. LVEF was measured by the 2-dimensional modified biplane Simpson method. The present study adopted a reduction of $\geq$ 15% in LVESV to define volumetric response to CRT, which has been widely accepted as the boundary between responders and non-responders [16–18].

## Preprocessing

Preprocessing of the MIT-BIH dataset and our own dataset is an essential step towards achieving accurate results when training machine learning models for ECG signal analysis. In this study, several preprocessing techniques were applied to the raw ECG data to improve the quality of the data and extract relevant features.

The raw ECG data underwent several preprocessing steps, including a high-pass filter to remove the baseline constant signal, and R-peak detection using a Chebyshev type I fourth-order filter and Shannon energy filter.[19,20] The ECG data were then segmented into 1.2 RR intervals.

Next, the 1D ECG signals representing HF were transformed into 2D time-frequency spectrograms using a short-time Fourier transform (STFT) with a Hamming window.[21,22] ECG signals are non-stationary data whose instantaneous frequency varies over time, and the properties of these changes cannot be fully described by using only frequency domain information. The STFT is an improved mathematical method derived from the discrete Fourier transform and used to explore the instantaneous frequency and amplitude of localized waves with time-varying characteristics. When analyzing a non-stationary signal, it is assumed to be approximately stationary within the duration of the temporal window of finite support.[23,24] The time-frequency spectrogram is given as follows:

$$X(\tau, w) = \int_{-\infty}^{\infty} x(t) w(t - \tau) e^{-jwt} \, dt$$

where $x(t)$ is the ECG signal, which is sampled at 360 Hz, and $w(t - \tau)$ is the Hanning window function with 512 window size that helps to smooth the signal at the edges of each time segment, reducing spectral leakage and improving the frequency resolution of the transform. The signal preprocessing is performed by the Python library Scipy [25]. A sample of some leads' spectrogram is shown in Figure 2.

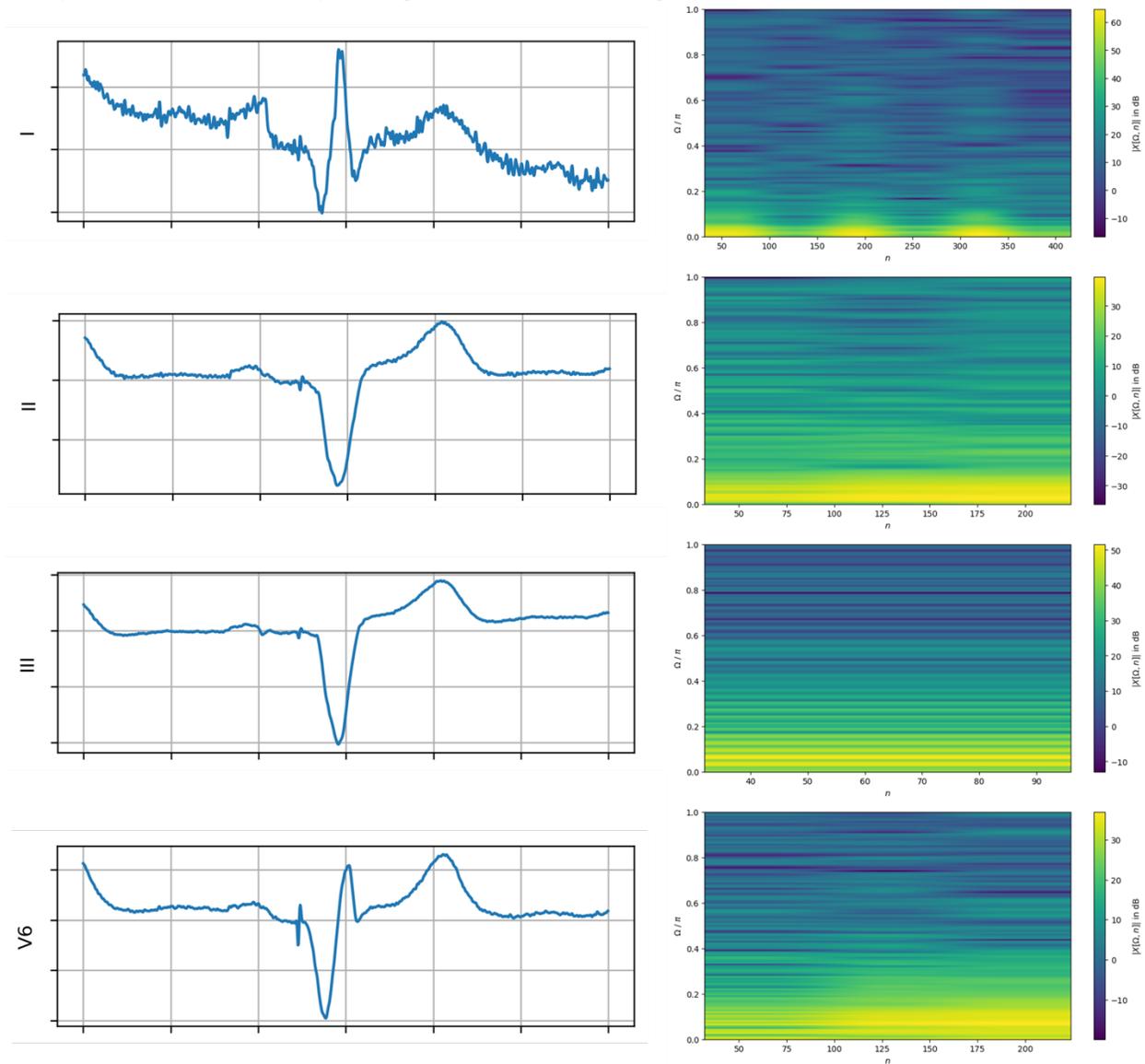

Figure 2. ECG spectrogram of sample data

### Resnet and Transfer Learning

This study used three CNN models, ResNet18, ResNet50, and ResNet101, to perform transfer learning on the preprocessed MIT-BIH dataset. The ResNet models are deep neural networks that use residual connections to address the problem of vanishing gradients during training [26].

ResNet is a deep neural network architecture that was introduced by He et al. in 2015 [26]. It is a variant of the traditional CNN that addresses the problem of vanishing gradients in deep networks. The ResNet architecture consists of residual blocks, enabling the network to learn a residual mapping instead of a direct one. The input to a ResNet block is a feature map x with dimensions $H \times W \times C$, where $H$ and $W$ are the spatial dimensions and $C$ is the number of channels. The output of the block is also a feature map y with the same dimensions.

The residual function $\mathcal{F}$ can be expressed as:

$$\mathcal{F}(x; \theta) = \mathcal{H}(x; \theta) - x$$

where $\theta$ are the learnable parameters of the residual function, and $\mathcal{H}$ is a set of convolutional layers followed by batch normalization and ReLU activation.

The output of the block is then computed as:

$$y = \sigma(\mathcal{F}(x; \theta) + x)$$

where $\sigma$ is an element-wise activation function (e.g., ReLU or sigmoid). This formulation allows the network to learn residual mappings that are easier to optimize than the original mappings.

Transfer learning is a powerful technique in deep learning that enables us to leverage pre-trained models to solve new tasks with limited data. In transfer learning, we start by pre-training a model on a large dataset, typically using a supervised learning approach. We then use this pre-trained model as a starting point for a new task with a different input and output domain. The pre-trained model is fine-tuned on the new task using a smaller dataset, which typically leads to better performance than training a new model from scratch. In transfer learning, the pre-trained model acts as a feature extractor, and the final layers of the model are modified to adapt it to the new task. We can also freeze some or all the layers in the pre-trained model to prevent overfitting on the new dataset. This approach allows us to achieve state-of-the-art performance on new tasks with limited labeled data.

## Results

The baseline characteristics are shown in Table 1. To evaluate the performance of a prediction model, it is important to use appropriate evaluation metrics. One commonly used metric is accuracy, which measures the percentage of correct predictions over the total number of predictions made. Sensitivity measures the proportion of true positives (i.e., correctly identified positive cases) among all actual positive cases, while specificity measures the proportion of true negatives (i.e., correctly identified negative cases) among all actual negative cases. Mathematically, these metrics can be defined as follows:

$$\text{Accuracy} = \frac{TP + TN}{TP + FP + TN + FN}$$

$$\text{Sensitivity} = \frac{TP}{TP + FN}$$

$$\text{Specificity} = \frac{TN}{TN + FP}$$

where $TP$ represents true positives, $FN$ represents false negatives, $TN$ represents true negatives, and $FP$ represents false positives. Sensitivity and specificity can provide insights into how well a model is able to detect positive and negative cases, respectively. A high sensitivity indicates that the model can identify most positive cases, while a high specificity indicates that the model can correctly identify most negative cases. However, these metrics can be influenced by the threshold for classifying predictions as positive or negative. It is important to choose an appropriate threshold that balances sensitivity and specificity for the specific application.

Table 1. Baseline characteristics of the enrolled patients

| Variable | All(n=71) | Response (N=46, 64.8%) | Non-response (n=25, 35.2%) | P value |
|---|---|---|---|---|
| ACEI/ARB | 58 (81.7%) | 36 (78.3%) | 22 (88%) | 0.318 |
| Age | 61.8 ± 12.6 | 61 ± 13.1 | 63.2 ± 11.7 | 0.487 |
| Gender | 56 (78.9%) | 34 (73.9%) | 22 (88.0%) | 0.170 |
| Height(cm) | 167.1 ± 7.1 | 167.0 ± 6.7 | 167.4 ± 8 | 0.832 |
| Weight(kg) | 67.8 ± 14.7 | 67.0 ± 13.6 | 69.4 ± 16.8 | 0.505 |
| CKD | 5 (7.0%) | 2 (4.3%) | 3 (12.0%) | 0.235 |
| DM | 15 (21.1%) | 6 (13.0%) | 9 (36.0%) | 0.024 |
| Hypertension | 33 (46.5%) | 22 (47.8%) | 11 (44.0%) | 0.762 |
| Smoking | 30 (42.3%) | 20 (43.5%) | 10 (40.0%) | 0.781 |
| Beta-blocker | 64 (90.1%) | 40 (87.0%) | 24 (96.0%) | 0.228 |
| Spironolactone | 61 (86%) | 38 (82.6%) | 23 (92.0%) | 0.284 |
| Digoxin | 14 (19.7%) | 8 (17.4%) | 6 (24.0%) | 0.511 |
| Diuretic | 61 (86.0%) | 40 (87.0%) | 21 (84.0%) | 0.737 |
| QRS duration | 172.0 ± 23.4 | 179.4 ± 19.8 | 158.4 ± 23.7 | 0.000 |
| LBBB | 57 (80.2%) | 44 (95.7%) | 13 (52.0%) | 0.000 |
| LVEF (%) | 26.9 ± 5.1 | 27.8 ± 5.1 | 25.2 ± 4.7 | 0.035 |
| LVEDV | 286.1 ± 85.2 | 272.9 ± 86.4 | 310.4 ± 78.9 | 0.076 |
| LVESV | 211.4 ± 74.5 | 199.1 ± 74.9 | 234.2 ± 69.6 | 0.057 |
| Scar | 26.3 ± 2.1 | 23.2 ± 11.2 | 32.0 ± 11.8 | 0.003 |

Data are expressed as mean ± SD or number (percentage)

The results presented in Table 2 were obtained through 5-fold cross-validation in our own ECG database. It can be observed that the prediction performance of various models using only the small ECG database is not satisfactory when compared to clinical guidelines, with the highest accuracy of 0.629, a sensitivity of 0.531, and a specificity of 0.732. However, when transfer learning is utilized to learn from a large-scale public database, significant improvements are observed in the performance of the models with an accuracy of 0.721, a sensitivity of 0.783, and a specificity of 0.792.

Table 2. Performance comparison of deep learning models

| Models | General methods | | | Pretrained methods (transfer learning) | | |
|---|---|---|---|---|---|---|
| | Accuracy | Sensitivity | Specificity | Accuracy | Sensitivity | Specificity |
| Guideline | 0.6 | 0.93 | 0.038 | | | |
| SVM | 0.58 ± 0.014 | 0.8 ± 0.016 | 0.333 ± 0.011 | | | |
| Random Forest | 0.58 ± 0.014 | 0.91 ± 0.017 | 0.125 ± 0.015 | | | |
| ResNet-18 | 0.613 ± 0.012 | 0.529 ± 0.016 | 0.71 ± 0.014 | 0.684 ± 0.012 | 0.6 ± 0.015 | 0.778 ± 0.018 |
| ResNet-50 | 0.612 ± 0.012 | 0.53 ± 0.015 | 0.73 ± 0.012 | **0.721 ± 0.015** | **0.783 ± 0.011** | **0.792 ± 0.017** |
| ResNet-101 | 0.629 ± 0.012 | 0.531 ± 0.016 | 0.732 ± 0.013 | 0.693 ± 0.015 | 0.674 ± 0.011 | 0.748 ± 0.013 |

## Discussion

In this study, we present a method for screening CRT patients based on deep learning techniques, demonstrating that pre-training from a large database of ECG arrhythmias and subsequently fine-tuning it on a small local database of CRT patients can significantly improve the performance of the target task, effectively reducing the inability to obtain knowledge of the ECG signal of the arrhythmia due to the small amount of data, and using this prior knowledge to help us to screen CRT patients quickly and efficiently. In the process of the proposed method, the time-domain ECG signal was transformed into a two-dimensional time-frequency ECG spectrum by a short-time Fourier transform. The resulting ECG spectrogram is used as the input to the proposed method. ECG arrhythmias were identified and classified using ResNet. The results show that the average accuracy of the ECG signal based on the 2D convolutional neural network can reach 72.1% for the prediction of CRT response. In addition, we did a series of comparison experiments to achieve the best classification performance with different parameter sets and structures of ResNet models. We found that the classifier based on the proposed 2D-ResNet50 model has the highest accuracy and the lowest loss when the learning rate is 0.001, and the batch size is 2000.

For QRS duration, current guidelines for CRT patient selection emphasize the importance of QRS duration as a criterion provided by the American Heart Association and American College of Cardiology [2,27]. Recent clinical trials have further highlighted the significance of QRS duration in CRT patient selection. These guidelines recommend that CRT implantation should be performed in patients who meet the class IA recommendation criteria[28], which include LVEF ⩽35%, QRS duration ⩾150ms, and the presence of

LBBB morphology. These guidelines recommend CRT implantation in patients with a LVEF of ≤35% if the QRS duration is ≥150ms and LBBB morphology is present. Studies such as CARE-HF[29], MIRACLE[30], REVERSE[31], and RAFT[32] have demonstrated that QRS duration is a powerful predictor of CRT outcomes, including mortality and morbidity. The findings from these trials have reinforced the importance of QRS duration as a reliable marker for CRT patient selection. However, The Echo-CRT trial studied patients with HFrEF and QRS duration <130 ms and found that CRT did not provide significant clinical benefit in this subgroup[33]. These findings suggest that patients with QRS duration below the current guideline threshold may not derive substantial benefit from CRT. And, compared to the 2013 guideline provided by the European Society of Cardiology, the lower QRS duration threshold was raised from 120ms to 130ms in 2021 guideline for HR patients in SR with LVEF≤35% and LBBB [28]. Despite its widespread use, QRS duration as a sole criterion for CRT patient selection has certain limitations. QRS duration primarily reflects electrical activation and may not fully capture the underlying mechanical dyssynchrony [7,34]. Some patients with narrow QRS complexes may still exhibit significant mechanical dyssynchrony, leading to suboptimal response to CRT[35,36] . Additionally, relying solely on QRS duration may exclude patients who could potentially benefit from CRT but have QRS durations below the threshold.

For QRS morphology, CARE-HF[29], MIRACLE[30], REVERSE[31], and RAFT[32] trials demonstrated that QRS duration was a powerful predictor of CRT outcomes (mortality and morbidity) compared to QRS morphology. Moreover, The MADIT-CRT[3,37] studies showed that even in non-LBBB patients, CRT can still provide benefit. These trials suggest that QRS morphology alone may not be sufficient to accurately predict response to CRT. While QRS morphology is an important criterion, it has certain limitations for CRT patient selection. QRS morphology primarily reflects electrical activation patterns and may not fully capture the underlying mechanical dyssynchrony in the ventricles. Some patients with different and heterogeneous electrical and mechanical activation patterns may exhibit similar QRS morphology, leading to a suboptimal response to CRT [38].

Recently, deep learning approaches have been widely used to improve the diagnosis of cardiovascular diseases with ECG. Attia et al. [39] proposed a CNN model to identify patients with ventricular dysfunction based on 12-lead ECG (AUC 0.93, accuracy 85.7%). In our work [40], we proposed an end-to-end ECG signal classification method based on a CNN model for the automatic identification of QRS morphology (five classes: normal beat, LBBB, RBBB, ventricular ectopic beat, and paced beat) using the MIT-BIH arrhythmia database [41]. Our CNN model achieved a classification accuracy of 0.9745, a sensitivity of 0.97, and an F1-score of 0.97 in identifying five classes recommended by the Association for Advancement of Medical Instrumentation.

Transfer learning has proven to be a valuable technique to handle the insufficient amount of annotated data that plagues trained classification models of ECG records, particularly for the detection of cardiac arrhythmias and abnormalities. Van [42] innovatively proposed transfer learning from the human ECG dataset to the equine ECG database for classifying four types: normal, premature ventricular contraction, premature atrial contraction, and noise. In the study by Weimann [43], different ResNet models pre-trained on the

Icentia11K5 data set with 11,000 patients, were fine-tuned on the PhysioNet/CinC Challenge 2017 data set consisting of 8528 labeled episodes to classify ECG signals into normal sinus rhythm, atrial fibrillation, and noise (too noisy to classify). And the ResNet-34v2 model in the pretraining task for beat classification achieved a performance of .794 ± .018 in the final test. Naz et al. [44] extracted deep features from different output layers of pre-trained AlexNet, VCG19, and Inception-v3 models; after selecting the best features, they used cubic support vector machine to perform the final classification on the MIT-BIH dataset, resulting in an accuracy of 97.6%.

In medical image analysis, the STFT is often used to transform ECG signals into 2D images for input into deep learning models. This allows the model to learn spatial patterns in the signal in addition to temporal patterns [45,46]. Moreover, it has been shown that without additional manual preprocessing of ECG signals, the accuracy of converting ECG signals into time-frequency spectrograms by short-time Fourier transform as input to 2D-CNN (99.00%) to predict the type of ECG arrhythmias is better than that of 1D-CNN (90.93%) [24].

## Conclusion

In this study, we propose an end-to-end ECG classification framework using a 2D CNN classifier. Using the STFT to transform 1D waveforms into 2D frequency-time spectrograms, our framework integrates a generalized pre-trained 2D CNN model for predicting whether a patient corresponds to CRT. The proposed approach outperforms existing clinical guidelines and popular machine learning models.